\documentclass[preprint,showpacs,preprintnumbers,amsmath,amssymb,superscriptaddress]{revtex4}

\usepackage{graphicx}
\usepackage{dcolumn}
\usepackage{bm}
\usepackage{pstricks}
\usepackage{epsfig}
\usepackage{psfrag}
\begin{document}


\title{Coulomb Oscillations of Indium-doped ZnO Nanowire Transistors in a Magnetic Field}

\author{Xiulai Xu}
\email{xx757@cam.ac.uk}

\affiliation {Hitachi Cambridge Laboratory, Hitachi Europe Ltd., JJ Thomson Avenue, Cambridge CB3 0HE, United Kingdom}

\author{Andrew C. Irvine}%
\affiliation{Microelectronics Research Centre, Cavendish
Laboratory, University of Cambridge, JJ Thomson Avenue, Cambridge CB3
0HE, United Kingdom}%

\author{Yang Yang}
\affiliation{Department of Engineering, University of Cambridge, 9 JJ Thomson Avenue, Cambridge CB3 0FA, United Kingdom}

\author{Xitian Zhang}%
\email{xtzhangzhang@hotmail.com}
\affiliation{Heilongjiang Key Laboratory for Advanced Functional Materials and Excited State Processes, School of Physics and Electronic Engineering, Harbin Normal University, Harbin, 150025, P.R. China}%

\author{David A. Williams}%
\affiliation{Hitachi Cambridge Laboratory, Hitachi Europe Ltd., JJ Thomson Avenue, Cambridge CB3 0HE, United Kingdom}%

\date{\today}

\begin{abstract}

We report on the observation of Coulomb oscillations from localized quantum dots superimposed on the normal hopping current in ZnO nanowire transistors. The Coulomb oscillations can be resolved up to 20 K. Positive anisotropic magnetoresistance has been observed due to the Lorentz force on the carrier motion. Magnetic field-induced tunneling barrier transparency results in an increase of oscillation amplitude with increasing magnetic field. The energy shift as a function of magnetic field indicates electron wavefunction modification in the quantum dots.

\end{abstract}

\pacs{72.80.Ey,73.23.Hk,73.63.Nm, 85.35.Gv,75.47.-m}




\maketitle
\section{Introduction}
Zinc oxide (ZnO) nanowires have received great interest for their potential applications in nanoelectronics and optoelectronics at room temperature
(for reviews, \cite{Ellmer2001,Meyer2004,Heo2004,Law2004,Ozgur2005}) because of the large direct band gap (3.37 eV at 300K) and the large exciton binding energy (60 meV).
Recently, many applications such as UV nano-laser, UV detector, gas sensor, field effect transistor,
solar cell and piezoelectric nanogenerator have been reported based on ZnO nanowires. The greatest barrier to ZnO being used in optoelectronic devices (for example,
light emitting diodes) is to realize stable and reproducible p-type doping \cite{Lyons2009}. Normally, ZnO is considered as a `native' n-doped material because of the
presence of intrinsic defects such as O vacancies (V$_{o}$) \cite{Ozgur2005,Xu2001,Carrasco2004}, Zn interstitials (Zn$_{i}$) \cite{Look1999} ,
 H interstitials \cite{Walle2000,Hofmann2002,Selim2007,Qiu2008} and Zn$_{i}$-V$_{o}$ complexes \cite{Look2005,Kim2009}. Among these defects, which one is dominant is still under debate.
Because of the uncertainty of the doping mechanism, investigations of the transport mechanism of ZnO nanowires are also challenging. Several transport mechanisms in a disordered system have been suggested for different temperature regimes \cite{Pepper1977}, such as thermal activation and hopping processes. For ZnO nanowires, different transport processes including thermal activation of electrons from the Fermi level to the conduction band, or to the impurity band \cite{Chiu2009}, variable range hopping (VRH) \cite{Ma2005} and nearest neighbor hopping (NNH) \cite{Chiu2009} have been reported. Recently, weak localization has been observed in indium-doped ZnO nanowires with a negative magnetoresistance \cite{Thompson2009}. However, to the best of our knowledge, Coulomb-blockade transport has not been reported in ZnO nanowire transistors.

Single electron tunneling via an island, the Coulomb blockade effect, has been investigated for more than 20 years \cite{Grabert}. Using the Coulomb interaction of electrons in quantum dots, single electron memory and single electron logic gates have been demonstrated \cite{Stone1998,Thelander2005}. In particular, single-charge measurement has been applied to demonstrate single-charge quantum bit readout \cite{Groman2005} and single spin readout \cite{Johnson2005} for quantum information processing. Single-electron effects also provide an opportunity to investigate the fundamental physics of single quantum dots. Coulomb blockade has been observed in different materials systems including Si, GaAs, carbon nanotubes and graphene, but not in transistors based on ZnO. This absence may be due to a number of factors: it is not easy to form an ohmic contact on ZnO because of its wide band gap, most research on ZnO is focused on achieving high temperature devices and, most importantly, Coulomb oscillations may smear out above a certain temperature range, which will be discussed in this work. Ma et al. \cite{Ma2005} observed a voltage gap of 0.8 V, which was attributed to the Coulomb gap. However, an electric-field induced current oscillation will be more convincing to demonstrate single-electron charging effect in the Coulomb-blockade regime \cite{Fulton1987,Scott1989}.

In this work, the conductance of ZnO nanowire transistors was characterized as a function of temperature. At low temperature, the main transport mechanism is hopping. In parallel with the hopping current, Coulomb oscillations from a localized single island or multiple islands in the nanowire were observed. With an applied magnetic field, a positive magnetoresistance was observed in the hopping transport regime. Coulomb peak height, however, increases with increasing magnetic field in spite of the magnetic field-induced confinement of the electron wavefunction. The energy shifts as a function of magnetic field are briefly discussed.

\section{Experimental details}

The indium-doped ZnO nanowires were synthesized using a vapor transport process. More details regarding the growth can be found in ref. \cite{Zhang2009,Wu2008}. The length of
the nanowires can be up to 30 $\mu$m and the diameter varies from 150 nm to 500 nm. The nanowires were transferred from the substrate and dispersed in isopropyl alcohol, and then spin coated onto a
n-type Si wafer covered with 200 nm of SiO$_{2}$. The position of the nanowire was identified using a scanning electron microscope aided by pre-patterned registration marks. The source and drain
electrodes (Ti/Au, 20 nm/100nm) were deposited on the nanowire using standard electron beam lithographic techniques and lift-off. The inset of Figure 1 shows a scanning electron microscope image of one of the devices, in which the source and drain are separated by around 3 $\mu$m. For the nanowire transistors in this work, the conductive substrate was used as the back gate. The electrical measurements were performed using a Heliox refrigerator with a temperature varying from 1.5 K to room temperature.

\section{Results and discussion}
\subsection{Temperature dependent conductivity}
Figure 1 shows the conductivity of the nanowire as a function of temperature. It can be seen that the conductivity decreases with decreasing temperature as expected for semiconducting behavior \cite{Chiu2009,Hu2009}. For a disordered semiconductor system, the temperature dependence of the conductivity is given by \cite{Pepper1977,Chiu2009}
\begin{equation}
  \sigma=\sigma_{1}exp(-E_{1}/k_{B}T)+\sigma_{2}exp(-E_{2}/k_{B}T)+\sigma_{3}exp(-E_{3}/k_{B}T)
\end{equation}
where $\sigma_{i}$ ($i=1,2,3$) are temperature-independent resistivity parameters. $E_{1}$ and $E_{2}$ are thermal activation energies for the excitations of electrons to the conduction band and impurity band, respectively. $E_{3}$ is average hopping energy at low temperature. With this equation, the temperature dependent conductivity can be well fitted as shown by the red curve in Figure 1. The fitted $E_{i}$ ($i=1,2,3$) are 62.0$\pm5$, 8.6$\pm0.5$ and 0.91$\pm0.2$ meV respectively, which are similar to what has been observed before. An $E_{3}$ of around 1 meV can be attributed the average hopping energy, as reported \cite{Chiu2009,Kumar2008}. Discussion of whether VRH or NNH dominates the transport in this temperature regime is beyond the scope of this work.

\subsection{Coulomb oscillations}
The $I-V$ characteristics of a transistor at different gate voltages at 1.5 K are shown in Figure 2 (a). It can be seen that the source-drain current ($I_{sd}$) almost linearly depends on the source-drain voltage ($V_{sd}$), indicating good ohmic contacts. $I_{sd}$ does not change very much with the gate voltage ($V_{g}$) from -20 V to +20 V, which means that the back gate does not sufficiently gate the whole wire. Furthermore, $I_{sd}$ does not increase monotonously with increasing $V_{g}$. In order to understand the gate-dependent current behavior, $I_{sd}$ as a function of $V_{g}$ for different $V_{sd}$ is plotted in Figure 2 (b). Clearly, $I_{sd}$ becomes increasingly negative with more negative $V_{sd}$. On top of that, the current is oscillating with $V_{g}$, and this is reproducible for different $V_{sd}$. The current oscillation amplitudes increase with $V_{sd}$ when $V_{sd}$ is small, then are smeared out at large $V_{sd}$. We ascribe the oscillations to the Coulomb oscillations of localized quantum dots in the wire \cite{Grabert}, which is due to the disordered potential within the wires \cite{Aravind2009}.

To observe the Coulomb oscillations clearly, differential conductance ($dI_{sd}/dV_{sd}$) as a function of gate bias is usually plotted. However, in our case, $I_{sd}$ has a large Ohmic current background, which increases linearly with $V_{sd}$. Therefore, plotting differential conductance as a function of gate bias is difficult to resolve the Coulomb oscillations. Here, differential transconductance ($dI_{sd}/dV_{g}$) is used instead \cite{Hergenrother1993,Zhuang1998}. The differential transconductance ($dI_{sd}/dV_{g}$) as a function of $V_{g}$ with $V_{sd}$ from -20 to 20 mV is shown in Figure 3 (a) at 1.5 K and Figure 3 (b) at 10 K respectively. Clear Coulomb oscillations can be observed. Because the localized quantum dots are randomly distributed in the wire and the shape might be irregular, the tunneling barriers between a quantum dot with two electrodes cannot be exactly same. This results in asymmetry of the oscillations for the positive and negative source-drain voltages, as shown in the Figure. The period of the oscillations also varies with gate voltage, which is due to the fact there is more than one quantum dot contributing to the current change. This can be well explained with multi-tunneling junction theory in single electron transistors \cite{Grabert}. The amplitudes of the oscillations decrease with increasing temperature because of phonon-assisted tunneling, and approach zero at around 20 K.

Figure 3 (c) shows the contour plot of the differential transconductance as a function of gate voltage from -14 to -7 V. For each current peak, the differential transconductance has a positive peak and a negative peak from the left side peak and the right side of the peak respectively. The Coulomb diamonds are sketched by dark grey lines. Considering the current peak spaces, the gate capacitance ($C_{g}$) is calculated to be about $0.05\sim 0.08$ aF. Extrapolating the diamond edges, the total capacitance ($C_{\Sigma}$) is estimated to be about $14.5 \sim 20$ aF and the charging energy is about $8 \sim 11$ meV. However, the charging energy calculated using the stability diagram could be over estimated, because the source-drain bias was not applied directly across the quantum dots for such a long nanowire. Alternatively, the charging energy can be estimated with the periodicity and the linewidth of the oscillation peaks. With a very small $V_{sd}$, the linewidth of single oscillation peak is about 3.5 $k_{B}$T \cite{Meirav}. The periodicity corresponds to the charging energy because the gate voltage tuning with a magnitude of the charging energy is required to observe next Coulomb peak \cite{Grabert}. To be able to observe the Coulomb oscillations, the charging energy should be larger than the linewidth, 3.5 $k_{B}$T. With a temperature of 20 K, the charging energy is around 6.1 meV, which is slightly smaller than that estimated from stability diagram.

In the case when Coulomb oscillations are induced by a single quantum dot, a regular Coulomb diamond should be produced. Figure 3(d) shows a contour plot of the differential transconductance when there is only quantum dot dominating the Coulomb oscillations at 4.2 K. Regular Coulomb oscillations can be clearly resolved. The charging energy for this case is around 4.0 $\pm$0.5 meV, calculated with both stability diagram and with the periodicity and the linewidth of the current oscillations at zero source-drain bias. If we assume the quantum dot is a sphere, a quantum dot diameter of around 86 nm can be deduced with a static dielectric constant of ZnO at 8.2.

\subsection{Coulomb oscillations in a magnetic field}
To confirm the observation of a quantum dot with few electrons, magnetic field-dependent transport spectra have been used to demonstrate the electron ground state and spin effects \cite{Tarucha1996,Ashoori1993}. In addition, ZnO is a very promising material for spintronics application, with a high Curie temperature \cite{Dietl2000,Liang2009}. Recently, large anisotropic magnetoresistance has been observed in GaMnAs in the coulomb blockade regime \cite{Wunderlich2006}. As discussed above, the transport current in our nanowire transistor has two parts; normal hopping current and tunneling current through localized quantum dots. To show magnetic dependence of the hopping current, the magnetoresistances as a function of magnetic field are plotted in Figure 4 (a). The gate voltages are selected at conductance minima, such that only hopping conductance is considered. Positive magnetoresistance is observed and increases quadratically. This positive magnetoresistance is caused by the current paths distorted in the magnetic field due to inhomogeneity. This can also be confirmed by the fact that the longtitudinal magnetoresistance is smaller than transverse one \cite{Huberman1987}.

In addition to the base current, the current amplitude of Coulomb oscillations reflects the tunneling probability and depends on the overlap of wavefunctions in the dots and in the contacts. Therefore, the conductance is sensitive to the wave function configuration within the dot \cite{Rokhinson2001}. Figure 4 (b) shows two oscillation peak (A and B, as shown in the inset) intensities as a function of magnetic field in different directions. The Coulomb peak height increase with increasing magnetic field in both configurations. For both peak A and B, magnetic field-induced oscillation current increase in the case for $B_{\perp}I$ is larger than that for $B_{\parallel}I$, especially at high magnetic fields. As the hopping channels are in parallel with quantum dot channel, one assumption for the peak increase could be due to that the reduced hopping current with magnetic field increases the visibility of the oscillation current; the larger the reduction of background hopping current the higher the Coulomb oscillation peak. However, this assumption can be eliminated by considering at other peaks (Peak C and D in Figure 4(c)). For peak D, the current increases are similar for both $B_{\perp}I$  and $B_{\parallel}I$. But for peak C, the oscillation current increase is actually larger for the case when $B_{\parallel}I$. We therefore ascribe the oscillation current increase to the tunnel barriers becoming more transparent, as has been observed in silicon devices \cite{Paul} and carbon nanotube devices \cite{Minot2004}. The magnetic field induces the redistributions of the wavefunction and the density of states of quantum dots and leads, which may result in conductance increase. Normally, magnetic field further confines the electron wavefunctions in quantum dots, resulting a less effective gating by the back gate. Comparing with Figure 3 (c), the peak spaces along the gate voltage are wider in the same device with a magnetic field at $B=10$ T (as shown in Figure 5). For instance, the width of central diamond (as marked) increases from 2.95 V to 3.50 V.

 The differential transconductance peaks for  $-20~V\leq V_{g} \leq 20~V$ as a function of $B_{\perp}$ and $B_{\parallel}$ are shown in Figure 6 (a) and (b) respectively. It can be seen that the peak position shifts are not sensitive to the direction of B. This enables us to conclude that peak shifts with magnetic field are dominated by spin effects \cite{Rokhinson2001}. Overall, the peak separations increase with magnetic field, which is attributed to magnetic field-induced additional confinement. That the intensity of the peaks increase with magnetic field is due to magnetic field-induced transparency as shown in Figure 4. The energy shift as a function of magnetic field could be due to a couple of reasons. Firstly, it could be due to the Zeeman effect. For ZnO, the Zeeman energy calculated from $g\mu_{B}B$, is about 1.1 meV for 10 T, where $g$ is the g-factor for ZnO and $\mu_{B}=e\hbar/2m_{e}$, the Bohr magneton. Secondly, the influence of the magnetic field on the energy levels of quantum dots can be due to a shell filling effect of the dot with a 2D harmonic confining potential \cite{Tarucha1996}, although no specified electron number states can be assigned here. Finally, it is possible that multi dots contribute to the current in the Coulomb blockade regime, and that the peak shift is induced by the different conduction paths between different dots when the magnetic field modifies the potentials between the dots and the leads. The intensity dependences of P15 and P16 in Figure 6 on magnetic field could be due to this.

\section{Conclusions}
In conclusion, we have demonstrated a single-electron charging effect of localized quantum dots in ZnO-based nanowire transistors. Clear Coulomb oscillations have been observed with a temperature up to 20K. The nanowire conductance decreases with magnetic field because of the Lorentz force on the electrical transport path. The conductance of the Coulomb peak, however, increases with magnetic field due to the reduced tunnel barrier. Magnetic field-dependent energy level shifting could be due to the Zeeman effect, the shell filling effect and magnetic field induced confinement. We believe that transport in ZnO nanowire transistors in the few-electron regime will be useful in understanding the nanowire transport and doping mechanisms, making ZnO nanowires more promising for applications in optoelectronics \cite{Law2004}, spintronics \cite{Dietl2000} and quantum information science \cite{Groman2005}.

\begin{acknowledgments}
We wish to thank Thierry Ferrus, Andrew Ferguson, Paul Chapman, Gareth Podd, Aleksey Andreev and Joerg Wunderlich for very helpful discussion and comments.
\end{acknowledgments}

\newpage

\begin{figure*}
Figure Captions:
\centering
\epsfig{file=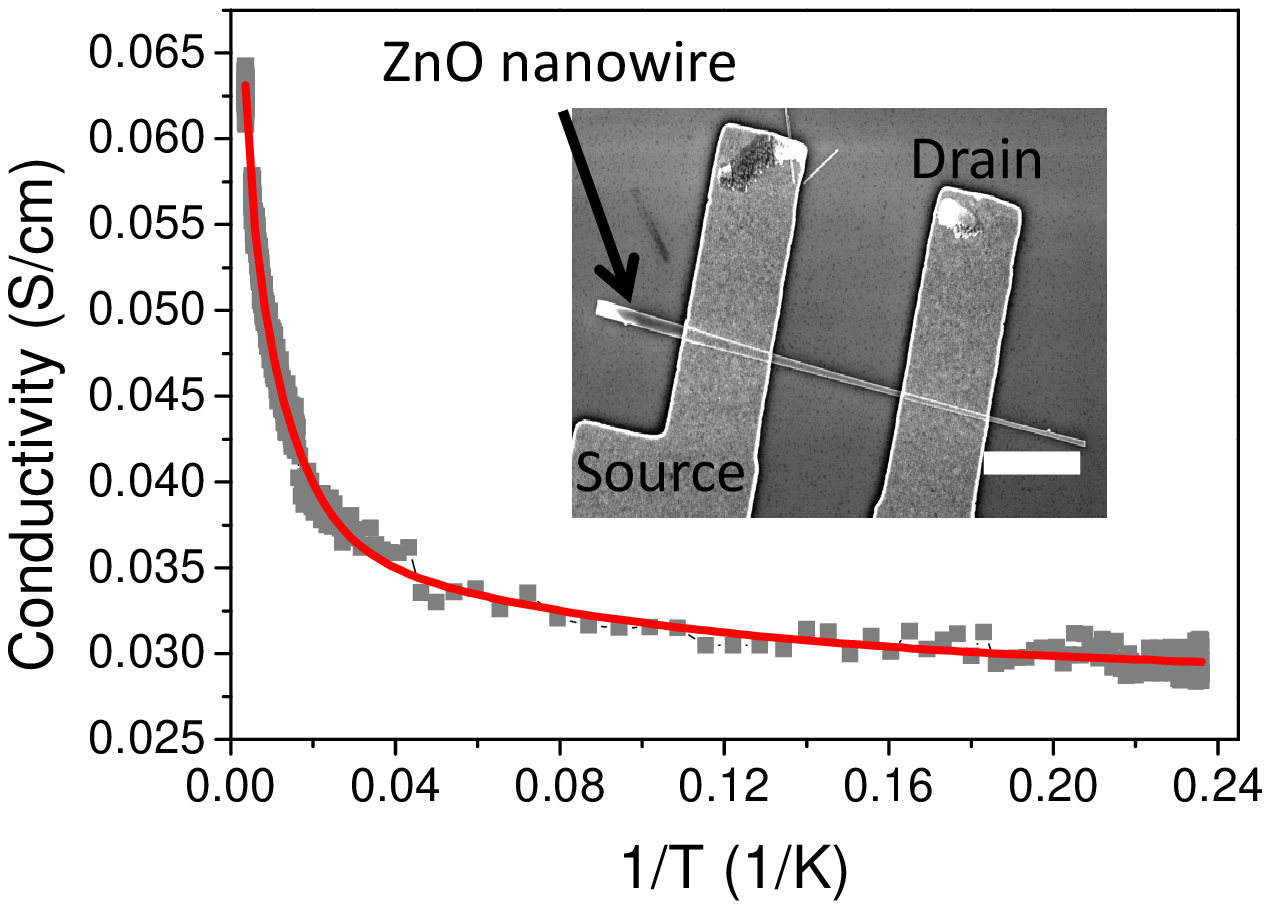,width=12cm,keepaspectratio}

\caption{\label{fig:figure1} Conductivity of a typical nanowire as a function of reciprocal temperature. Inset: A SEM image of a nanowire transistor. The white scale bar is 2 $\mu$m. The electrode width of 2 $\mu$m was used to improve the Ohmic contact.}

\end{figure*}

\begin{figure*}

\centering

\epsfig{file=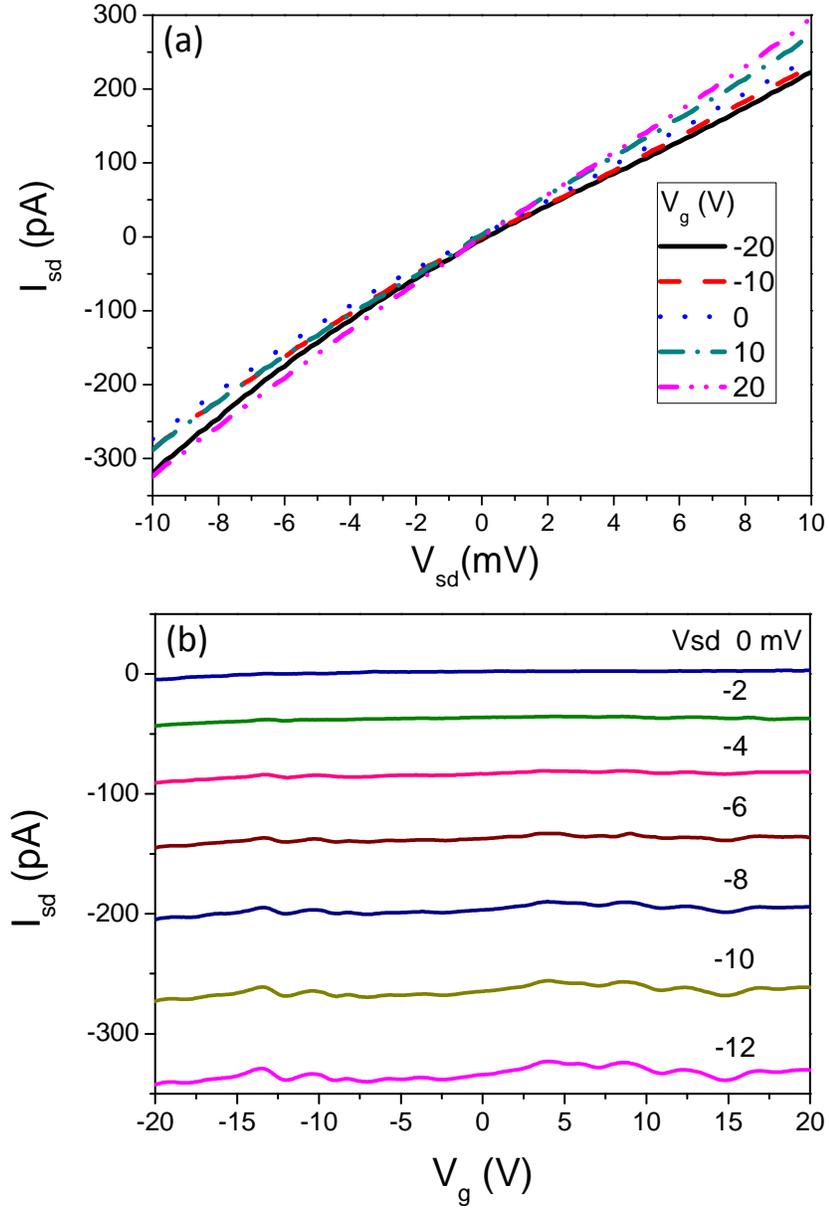,width=12cm,keepaspectratio}

\caption{\label{fig:figure2} (a) A typical I-V curve of a transistor with different back gate voltages. (b) Source-drain current as a function of gate voltage at different source-drain voltages.}

\end{figure*}

\begin{figure*}

\centering

\epsfig{file=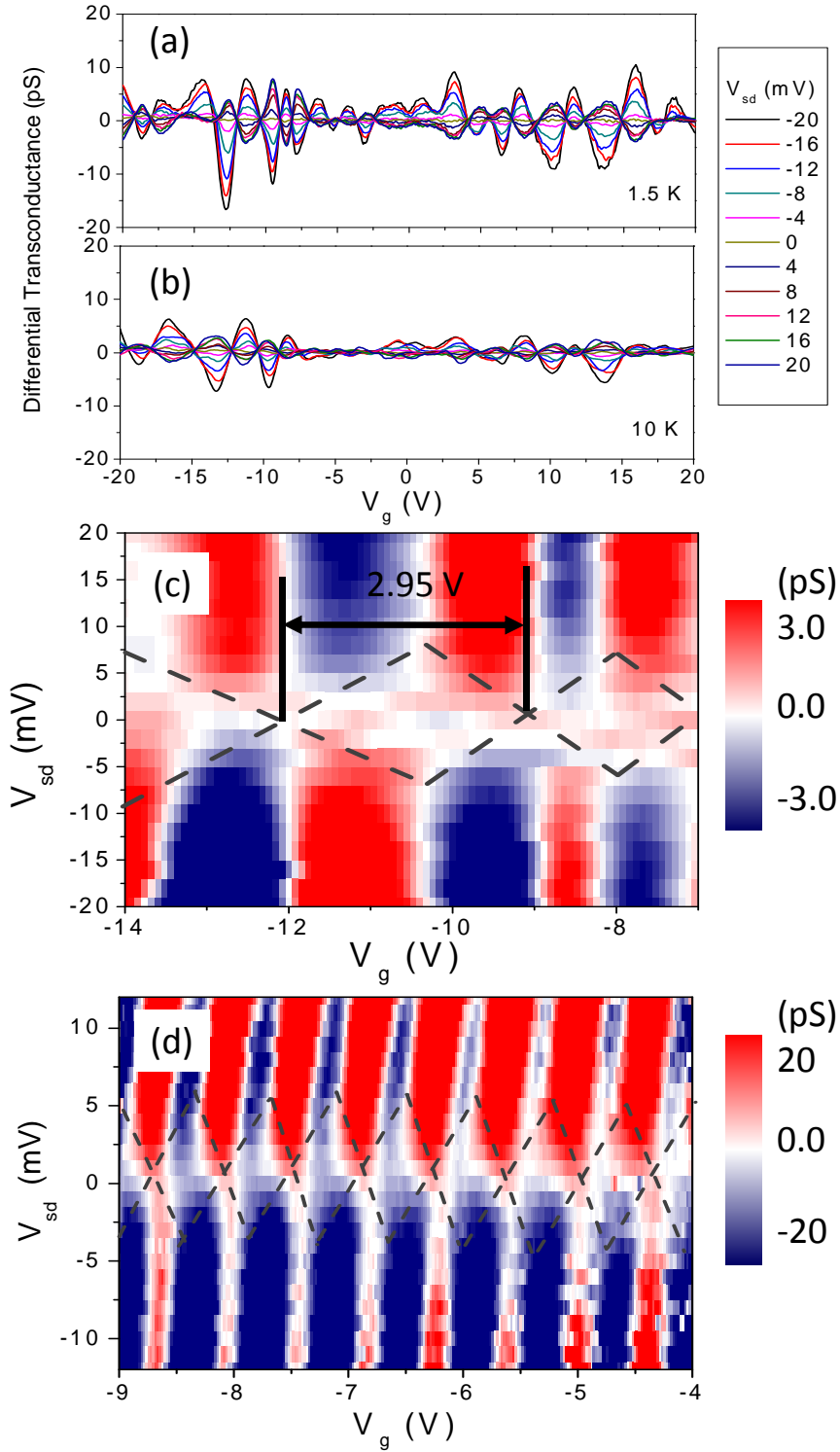,width=12cm,keepaspectratio}

\caption{\label{fig:figure3} (a) and (b) are differential transconductance as a function of gate voltage at 1.5 K and 10 K respectively. (c) Contour plot of the differential transconductance for a multi-tunneling junction device. The dashed dark grey lines are highlighting the Coulomb diamonds. (d) Coulomb diamonds of a transistor when a single island dominates the Coulomb oscillations. }

\end{figure*}

\begin{figure*}

\centering

\epsfig{file=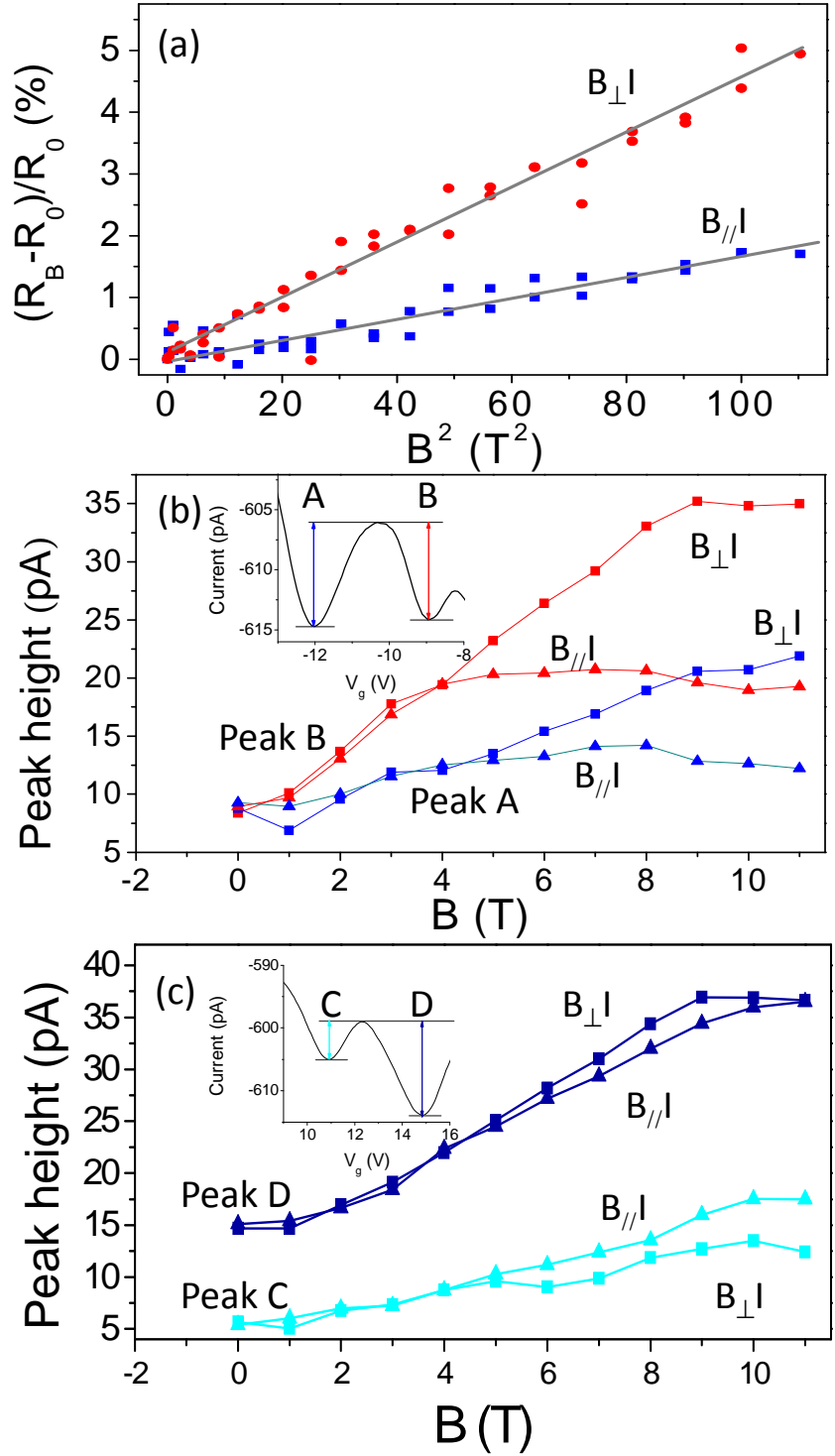,width=12cm,keepaspectratio}

\caption{\label{fig:figure4} (a) Magnetoresistance as a function of $B^{2}$ in configurations of $B_{\parallel}I $ and $B_{\perp }I$.  (b) and (c) Coulomb peak height as a function of magnetic field in $B_{\perp }I$ (solid squares)and $B_{\parallel}I $ configurations (solid triangles). The peaks are shown in the insets with a $V_{sd}$ at -20 mV.  }

\end{figure*}

\begin{figure*}

\centering

\epsfig{file=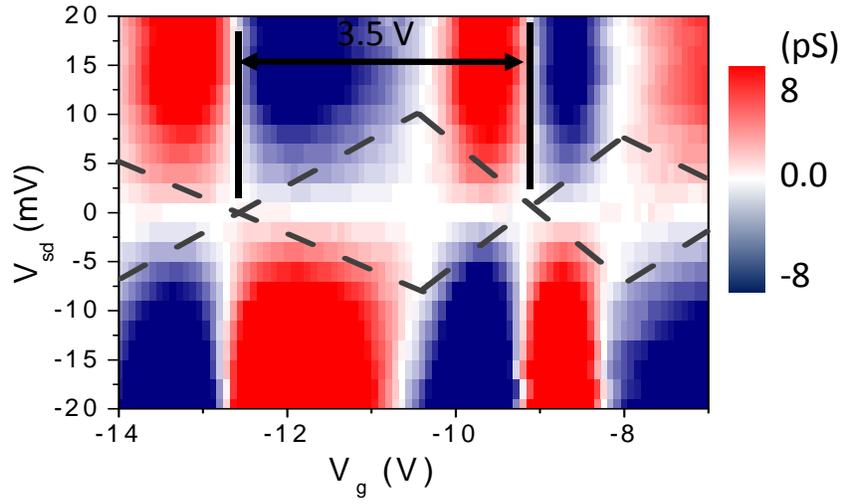,width=12cm,keepaspectratio}

\caption{\label{fig:figure5} Contour plot of the differential transconductance for the device in Figure 3 (c) at 1.5 K with $B=10$ T. The dashed dark gray lines are used to guide the eye. }

\end{figure*}

\begin{figure*}

\centering

\epsfig{file=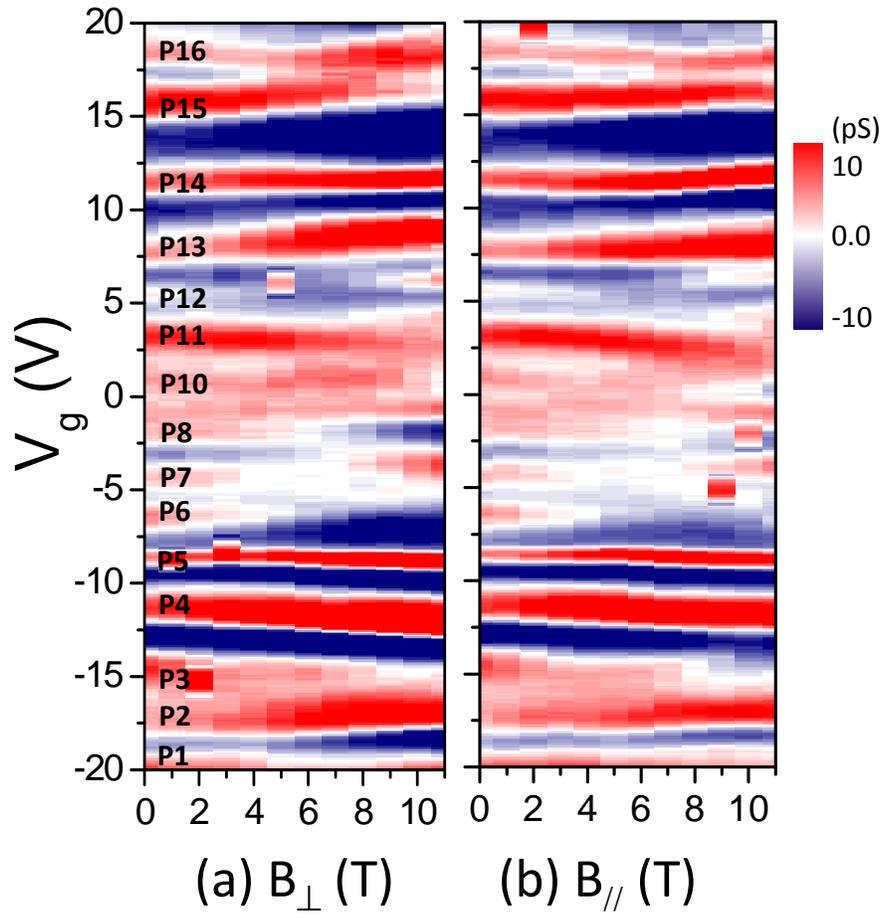,width=12cm,keepaspectratio}

\caption{\label{fig:figure5} Contour plots of differential transconductance as a function of both magnetic field and gate bias for  $B_{\perp }I $ (a) and $B_{\parallel}I $ (b). The peaks are labeled in (a).  }
\end{figure*}

\end{document}